\documentstyle[12pt]{article}
\begin{document}

\begin{flushright}{\bf\underline{
Revised Nov/1997}}\end{flushright}

\vspace{2cm}
\begin{center}
{\huge \bf The Ring Division Self Duality}

\vspace{3.5cm}

{\large \bf Khaled Abdel-Khalek
\footnote{Work supported by an
ICSC--World Laboratory scholarship.\\
 e-mail : khaled@le.infn.it}
}
\vspace{.4cm}

\end{center}

\vspace{3.5cm}

\begin{abstract}
We present a simple construction of the instantonic type equation
over octonions where its similarities and differences with
the quaternionic case are very clear. We use the unified language
of Clifford Algebra.
We argue that our approach is the pure algebraic formulation
of the geometric based soft Lie algebra.
The topological
criteria for the stability of our solution is given
explicitly to establish its solitonic property.
Many beautiful
features of the parallelizable 
ring division spheres and Absolute Parallelism (AP) reveal their presence
in our formulation.
\end{abstract}

\newpage

\vspace{0.5cm}
\section{Introduction}

During this century, physics has evolved a great deal. From general
relativity to quantum fields and finally by Yang-Mills gauge fields,
our mathematical tools have been amplified. It seems that our world has
signed an agreement with Dirac's ideology ``One can generalize his physics
by generalizing his mathematics''. For example, we know that our 
non-abelian gauge fields without any coupling to matter have a very rich
topological structure. Instantons represent a remarkable bridge between
modern mathematics  and  - eventually - physics.
The  extension of the 4 dimensional
Euclidean Yang-Mills
instanton \cite{bpst} to higher dimensions  
was formulated by different groups
\cite{gd1,gd,fub,WA,IP}. Recently, such topics appeared to the surface
with  important application in string solitons
\cite{duf1,pp3,gun1,Hst,duf2,a1,a2,a3,a4,a5,a6,a7}.
 
In this article, we would like
to give a new  construction for an octonionic
instantonic type equation
where the parallelism between such solutions in different dimensions
is very transparent. Also, the topological stability is written
explicitly to establish the solitonic, not just the integrability, 
property of such a model. 

We use the following notations: $dim = 2^n$, roman letters
$(a,b, ...)$ run from
1 to (dim - 1) whereas Greek letters $(\alpha,\beta ...)$ 
run from 0 to (dim - 1). We denote quaternions and octonions
by \cal{Q} and \cal{O} respectively.
The paper is organized as follows: in the next section,
we give a brief review of some geometric properties of
the ring division algebra then in section 3., we define a Clifford self
duality which enable us to represent, in section 4., a novel construction
of the eight dimensional instantons where we also
compute its ``torsionful cohomology group'', unfortunatly,
our solution does not satisfy the YM equation of motion
but
\begin{equation}
3 \partial_\mu F_{\mu \nu}  + [ A_\mu, F_{\mu \nu }] = 0.
\end{equation}
Then in section 5. we discuss the possibility of
extending this formulation to higher dimensions.
Lastly in  section 6.
we spell out the message of this paper.

\section{The AP Structure}

We have first to review some of the beautiful
algebraic, geometric and topological properties of the ring division
algebra.  We will see that different branches of mathematics must be used
together to give the correct physical picture. Using algebra alone and
writing down a multiplication table is not at all sufficient. 

One of the most important application of the ring division
algebra in modern mathematics is its connection with the
problem of finding the number of vector fields over any sphere.
A problem that had been solved completely by Adams\cite{adams} and then
the solution had been simplified and reformulated by different
great mathematicians. Following Husemoller \cite{husm}, one can simply
say
that the number of allowed vector fields N-1 that parallelize - i.e
the needed number of linearly independent vector fields, that
exist globally,
and form a basis of - a sphere
is related, on the one hand, to the ring division algebras
which single
out just three spheres $S^1, S^3$ and lastly $S^7$.
Whereas on the other hand, these  N-1 vector fields
are related to a  Cliff(0,N-1) structure
expressed as $N\times N$ real matrices. Again, which is only possible
in 1,3 and 7 dimensions. Leading to N=1,2,4,8 the dimensions of
the different ring division algebra.
In \cite{my1,my2}, We have shown how to construct  a Cliff(0,N-1)
directly from the ring division algebra
where they are simply given by
\begin{equation}
(E_i)_{\alpha\beta} = \delta_{i\alpha} \delta_{\beta 0} - \delta_{i \beta}
\delta_{\alpha 0} + f_{i\alpha\beta} ,
\end{equation}
$f_{ijk}$ is the ring division structure constant,
then we have 
\begin{equation}
\{ E_i, E_j \} = -2 \delta_{ij}  .
\end{equation}
Also note that 
\begin{equation}
(E_i)^T = - E_i , \quad \quad (E_i)^2 = - 1.
\end{equation}
At the level of quaternions, we have a complete
algebraic isomorphism
\begin{equation}
E_i E_j = -\delta_{ij} + f_{ijk} E_k ,
\end{equation}
whereas for octonions, one finds
\begin{eqnarray}\label{as1}
E_i  E_j  =  -\delta_{ij} + f_{ijk}
E_k -  [ E_i, 1|E_j ],
\end{eqnarray}
we use Rotelli's notation\cite{rot} where right action is denoted
by $1|E_i$.

But since the $S^7$ sphere is parallelized by octonions
or by the above Cliff(0,7) then they are geometrically  isomorphic
and this algebraic difference is just a reflection of comparing
an associative algebra to a non-associative one.
To be more convincing, let's review briefly this
Absolute Parallelism (AP) structure i.e how our Cliff(0,dim-1)
parameterize the different ring division $S^{dim-1}$ spheres.
AP spaces are non-trivial torsionfull manifolds. They  have a very
important characteristic: The vanishing of the  parallelizable
torsionfull  connection i.e. the covariant derivative will be reduced
to the standard derivative \cite{wolf,clarck}
\begin{equation}
D_\mu A_\nu = \partial_\mu A_\nu - \Gamma_{\mu \nu}^{\alpha} A_\alpha
 = \partial_\mu A_\nu.
\end{equation}
We are going to demonstrate how this happens for the ring division
spheres.
{\bf For quaternions or octonions,
we can use either of the $e_i$ or the $E_i$ sets. The important thing is
having a Cliff(0,3) ($4\times 4$ matrices) and a Cliff(0,7)
($8\times 8$ matrices) structure.}

Our line of attack is simply finding the parallelizable
coordinates frame where the metric will have a flat basis.
Working over $R^{dim}$, we have the following metric
\begin{equation}
ds^2 = dx_\mu dx^\mu ,
\end{equation}
embedding the $S^{dim-1}$ spheres amounts to impose
the condition
\begin{equation}
x_\mu x^\mu = R^2 ,
\end{equation}
which induces the Cartesian metric
\begin{equation}
ds^2 = \left(  \delta_{mn} +
{\frac{y^m y^n}{R^2 - y^a y^a}} \right) dy^m dy^n .
\end{equation}

To find an easy way to connect the spherical to the
Cartesian coordinates, that is the business of the Cliff(0,dim-1).
The complex case $n=1$ is well known, so starting
with $S^3$,
\begin{eqnarray}
q &=& 
x_0 + x_1 E_1 + x_2 E_2 + x_3 E_3 \nonumber\\
&=& |R| \exp (E_1 \theta_1 + E_2 \theta_2 + E_3 \theta_3 )
 .\end{eqnarray}
What did we gain? It should be clear that with
this special choice of $x_\mu$ we can introduce simply
the following spherical metric
\begin{eqnarray}
ds^2 &=& dq d\bar{q} = - R^2 E_i E_j d\theta_i d\theta_j , \\
&=& R^2 \delta_{ij} d\theta_i d\theta_j .
\end{eqnarray}

For octonions, everything can be done without modifications
\begin{eqnarray}
O &=& 
x_0 + x_1 E_1 + \ldots + x_7 E_7 \\
&=& |R| \exp (E_1 \theta_1 + \ldots + E_7 \theta_7 )
 , \\
\end{eqnarray}
Leading to 
\begin{eqnarray}
ds^2 &=& dO d\bar{O} = - R^2 e_i e_j d\theta_i d\theta_j , \\
&=& R^2 \delta_{ij} d\theta_i d\theta_j .
\end{eqnarray}

Going to higher dimensions, it is impossible to repeat such
constructions. According to {\bf the standard} classification
of Clifford Algebras\cite{abs}, Cliff(0,15) is represented by
$128\times 128$ matrices.

\section{The Self Duality}

To start, the 4 dimensional Yang-Mills self duality is
\begin{equation}\label{a1} 
F = \pm ^{*}F , 
\end{equation} or
 in terms of components 
\begin{equation}\label{a2} 
F_{\alpha\beta} = 
\pm{\frac{1}{2}} \epsilon_{\alpha\beta\mu\nu} 
F_{\mu \nu}. 
\end{equation} 
where $\epsilon_{\alpha\beta\mu\nu}$ is the Levi-Civita tensor. The possible
generalization of this self duality is proposed to be
\begin{equation}\label{a3} 
F_{\alpha\beta} = {\frac{1}{2}}
\eta_{\alpha\beta\mu\nu} F_{\mu \nu} , 
\end{equation} 
where
$\eta_{\alpha\beta\mu\nu}$ 
is a fourth rank tensor. 
Generally (\ref{a3}) is not $so(dim)$ invariant for
$dim>4$ since a generic $\eta_{\alpha\beta\mu\nu}$
invariant under the action of $so(dim)$ in arbitrary dimensions is not
available but it can be invariant with respect to a submanifold of
$so(dim)$. Here, we would like to fix this submanifold as $S^7$ and
present a simple method for determining $\eta$ explicitly. 

For octonions,
in contrast to quaternions,
{\bf our left and right actions do not commute which makes $S^7$ a
non-group manifold}  contrary to $S^3$. Simply because $S^7$ possess a
varying torsion whereas $S^3$ has a constant torsion
T(X,Y) equal to $- 2f_{ijk}$ since for AP spaces
\begin{equation}
\nabla_X Y = 0
\end{equation}
leads to 
\begin{equation}
T(X,Y) = \nabla_X Y - \nabla_Y X - [ X,Y ] = - [ X , Y] .
\end{equation}
In \cite{sp,ced1,brink,berk1,berk2}, using geometric tools,
the notion of soft Lie algebra has
been introduced where we no longer have structure constants but  structure
functions defined by our vector fields and they vary over $S^7$.
At the level of the algebra such a varying torsion can be seen
as follows\footnote{Eqn. (\ref{as3}) is similar to the
starting point for the
Knizhnik-Bershadsky arbitrary N superconformal algebra in two
dimensions\cite{kniz,ber}}:
\begin{eqnarray}\label{as3}
[E_i , E_j]  &=&  2f_{ijk}
E_k -  2[ E_i, 1|E_j ] ,\\
&=& 2(f_{ijk} 1_{8\times 8} + [E_i, 1|E_j] E_k ) E_k , \label{lop}\\
&=& 2\rho_{ijk} E_k  .
\end{eqnarray}
So our structure constants become matrices\footnote{There is no
summation in the second term of (\ref{lop}) to avoid an ugly
$1/7$ factor.} and our torsion
is
\begin{equation}\label{tor}
T(X,Y) = - 2\rho_{ijk}.
\end{equation}

Now to build a solitonic model, the idea is simply to gauge the $S^7$
and to compactify our Euclidean space from ${\cal R}^8$ to $S^8$
which is a non-trivial manifold and we should cover it by at least two
patches (e.g north and south semi-hyper-spheres) then we have to introduce
two different gauge fields which are equivalent up to a gauge transformation
in the overlapping region (the equator is $S^7$), so the gauge
transformations  define a map from the equator to the gauge algebra~:
\begin{equation}\label{map}
 S^7 \longrightarrow S^7 , \end{equation}
 we have\cite{tabl1},
 \begin{equation}\label{lll} \pi_{S^7}\left( S^7 \right) = Z ,
\end{equation}
establishing that our solutions will represent distinct
cohomology classes which is the required guarantee for the stability of our
instanton i.e our model is not just exactly solvable
 - integrable - but of solitonic type.  In the next section, we are going
 to prove
 that our solution satisfies (\ref{lll}).

The critical point for the self-duality condition is the natural existence
of the third rank antisymmetric tensor $\rho $ which
is important to determine
the fourth rank tensor of (\ref{a3}). Adding a zero index to extend
$\rho_{ijk}$ from $R^7$ to $R^8$, we
define
\begin{equation}\label{sel}
\eta_{0ijk} = \rho_{ijk} ,
\end{equation}
and zero elsewhere which enables us to introduce the Clifford self duality
condition for a 2-form as
\begin{eqnarray}\label{star}
F = \pm{^\star}F 
&\Longleftrightarrow& 
F_{0\mu} = \pm  {\frac{1}{2}} \eta_{0 \mu \alpha \beta } F_{\alpha \beta} , \\
&\Longleftrightarrow& F_{0i} = \pm {\frac{1}{2}} \eta_{0ijk} F_{jk} , \\
&\Longleftrightarrow& F_{0i} = \pm {\frac{1}{2}} \rho_{ijk} F_{jk}.
\end{eqnarray}
 This choice mimicks the 4-dimensional
electromagnetic duality
\begin{equation}
F_{0i} = - \epsilon_{ijk} F_{jk}  ,
\end{equation}
in contrast to the condition used in
the previous octonionic formulations where
$\eta_{abcd} = \epsilon_{abcdijk}
f_{ijk}$ which is very specific to octonions.
Also, it is evident that our condition
in 4 dimensions reduces to the standard one.

With the notation (\ref{sel}), $E_i$ is given by
\begin{equation}
(E_i)_{\alpha\beta} = \delta_{i\alpha} \delta_{\beta 0} - \delta_{i \beta}
\delta_{\alpha 0} + \eta_{0i\alpha\beta } ,
\end{equation} 
in complete agreement with 't~Hooft notation's \cite{hoft} leading 
to  
\begin{eqnarray}\label{as2}
[ E_i , E_j ] = 2\eta_{0ijk} E_k - 2 [ E_i, 1|E_j ] .
\end{eqnarray}

We have preferred to work explicitly with octonions but the same
construction holds equally well for any Lie group $\geq$  $so(7)$. It is just an
embedding problem which can be solved using the following facts :
\begin{equation}
so(7) \sim \{ \frac{1}{4} [E_i,E_j] \} , \quad
so(8) \sim \{ \frac{1}{4} [E_i,E_j] , \frac{1}{2} E_k \} ,
\end{equation}
and
\begin{equation}
\pi_7(so(7)) = Z , \quad   \pi_7(so(8)) = Z \bigoplus Z.
\end{equation}

\section{The 't~Hooft Solution}

An ansatz for the 't~Hooft \cite{hoft}   like
solution 
can be done but let's first construct a self-dual
basis.
Anyone who has the least knowledge about instantons
knows quite well the utility of such tensors and how they may be
used as a shortcut for many calculations.

Using octonions, we define
\begin{equation}
E_\mu \equiv ( 1 , E_i ) \quad , \quad \bar{E}_\mu \equiv ( 1 , - E_i ),
\end{equation}
and
\begin{equation}
\vartheta_{\mu\nu} =  {\frac{1}{2}}
( \bar{E}_\mu E_\nu - \bar{E}_\nu E_\mu ).
\end{equation}                                  
which is Clifford self Dual
\begin{equation}
\star \vartheta = \vartheta ,
\end{equation}
i.e
\begin{equation}
\star \vartheta_{\alpha \beta} = {\frac{1}{2}} \eta_{\alpha \beta \mu \nu }
\vartheta_{\mu \nu} = \vartheta_{\mu \nu}.
\end{equation}

We can now construct our solution.
Consider an $S^7$ element
\begin{equation}\label{gg}
g(x) = \frac{E_\mu x^\mu}{x^2} ,
\end{equation}
mimicking the quaternionic case, our self-dual gauge
is
\begin{equation}\label{sol}
A_\mu(x) = \frac{x^2 }{x^2  + \lambda^2}
g^{-1}(x) \partial_\mu g(x) = -
\frac{\vartheta_{\mu \nu} x^\nu}{\lambda^2 + x^2}
\end{equation}
leading to
\begin{equation}
F_{\mu\nu} =
\frac{\vartheta_{\mu \nu} 2 \lambda^2}{(\lambda^2 + x^2)^2} ,
\end{equation}
Which is Clifford self-dual
\begin{equation}
F_{\mu\nu} = {\frac{1}{2}} \eta_{\mu\nu\alpha\beta} F_{\alpha\beta}.
\end{equation}
An anti-self-dual can be done easily  using an anti-self-dual basis.

Now, the important point, we have to compute our ``torsionfull
cohomology group''. We know that for a quaternionic instanton
it is simply $\pi_3(S^3)$ given by (up to a normalization constant)
\begin{equation}
\int_{S^3}{I_1(\alpha , \beta , \gamma )} = \int_{S^3}{Tr(\epsilon_{\alpha\beta \gamma}
g^{-1}\partial_\alpha  g\ 
g^{-1}\partial_\beta  g\ 
g^{-1}\partial_\gamma  g)}
\end{equation}
by explicit calculation  of any one of these elements, yields
\begin{eqnarray}
I_1(1,2,3) &=& I_1(1,3,2) = I_1(2,1,3) = I_1(2,3,1) =
I_1(3,1,2) = I_1(3,2,1) \nonumber\\
&=& - \frac{4 x_0}{x^4}.
\end{eqnarray}
For octonions
\begin{equation}
\int_{S^7}{I_2(\alpha , \beta , \gamma )} = \int_{S^7}{Tr(\rho_{\alpha\beta \gamma}
g^{-1}\partial_\alpha  g\ 
g^{-1}\partial_\beta  g\ 
g^{-1}\partial_\gamma  g)}
\end{equation}
again any of these elements is
\begin{eqnarray}
I_2(1,2,3) &=& I_2(1,4,5) = I_2(2,4,6) = I_2(3,4,7) =
 \dots = {\mbox{all the possible symmetrization}} \nonumber\\
&=& - \frac{8 x_0}{x^4}.
\end{eqnarray}
It is clear that - apart from a normalization constant -
our $I_2$ is an element of $\pi_7(S^7)$.

Generally the 't~Hooft like solution is
\begin{eqnarray}
A_\mu =
- \frac{\vartheta_{\mu \nu} (x - y)^\nu}{\lambda^2 + (x - y)^2}
\end{eqnarray}
where $y$  are eight free parameters because of translation
invariance so our solution has 9k free parameters.

Having the topological stability criteria, our n-solitons
can never decay either to the trivial or any other m-solitons.
A generic (9k +8) n-solitons solution can also be written without
problems but 
a {\underline ``preliminary''} calculation of our moduli space, using
the methods developed
in this paper and some other Clifford Bundle techniques, indicates
that the dimension of the moduli space is 16 k - 7 so a twistor
construction similar to the ADHM solution\cite{ADHM} is needed
\footnote{Actually, this 16 k -7 can be counted directly from
the number of free parameters involved in a twistor like methods}.
We will return to this point elsewhere.

Now, problems start. In 4 dimensions, Instantons have
a very clear important meaning.
They satisfy the Yang-Mills equation of motion and represent the
absolute minima of our non-abelian gauge fields. Working with
quaternions $E_\mu$ ($\mu=1 \ldots 4$, (the $\times 4$ quaternionic matrices
can be found in \cite{my1}),
\begin{equation}
A_{\mu} = \frac{-\vartheta _{\mu \nu} x^\nu  }{\lambda ^2+ x^2}
\end{equation}
which satisfy
\begin{equation}
D_{\mu } F_{\mu \nu } = \partial_{\mu} F_{\mu \nu } + [A_{\mu },F_{\mu \nu}]
= 0 .
\end{equation}

Going to octonions ($\mu = 1 \ldots 8$), we find
that our solution (\ref{sol}) does not satisfy the Y.M. equation of motion.
\begin{equation}
D_{\mu }F_{\mu \nu } = \partial_{\mu}F_{\mu \nu } + [ A_{\mu },F_{\mu \nu} ]
\neq  0 .
\end{equation}
It comes too close
\begin{equation}
3  \partial_{\mu} F_{\mu \nu } + [A_{\mu },F_{\mu \nu}]  = 0,
\end{equation}
but in physics there is no difference between 3 or $10^{10}$.
Our solution fail to mimick exactly the 4 dimensional case.
In fact from the start, and contrary to the quaternionic
case, our Clifford self duality does not imply the Bianchi identities.

But only for $\lambda \longrightarrow 0$, the 't~Hooft like solution in
8 dimensions satisfy both of the Clifford self duality and
the YM eqn. of motion.
To this point, omitting such constraint, one should change the
form of the solution, for example, a pure gauge connection
\begin{eqnarray}
A_\mu = g^{-1} \partial_{\mu } g &\Longrightarrow&
F_{\mu\nu} = \star F_{\mu \nu} ,\nonumber \\
&\Longrightarrow& D_{\mu} F_{\mu\nu} = 0.
\end{eqnarray}
But, at least for the time being, we don't know yet the possibility
of such solution.
Recently, the authors of \cite{duf2} complained about the self duality of the
Fubini-Nicolai solution \cite{fub}. So, the 8 dimensional case is still open.
In summary, we solved a new 8 dimensions self duality, that is reducible to
4 dimensional case, but our solution, only in certain limit,
satisfy  the Yang-Mills
equation.
Of course the four dimensional case is more powerful
because the self duality is related directly to the Bianchi identity
which does not hold in higher dimensions.

\section{Higher Dimensions}

Going to higher dimensions, we define a hexagonions (${\cal X}$)
as
\begin{eqnarray}\label{ffff}
{\cal X} &=& {\cal O}_1 + {\cal O}_2 e_8 \\
&=& x_0 e_0 + \ldots + x_{16} e_{16}.
\end{eqnarray}
and
\begin{equation}
e_i e_j = -\delta_{ij} + f_{ijk} e_k.
\end{equation}
Now, we have to find
a suitable
form of $f_{ijk}$. Recalling how this structure constant 
is written for 
octonions
\begin{eqnarray}\label{a4}
O &=& {\cal Q}_1 + {\cal Q}_2 e_4 \\
&=& x_0 e_0 + \ldots  + x_7 e_7 ,
\end{eqnarray}
from $\cal Q$, we have already chosen $e_1e_2 =e_3$ and
from the decomposition 
of (\ref{a4}), we set $e_1e_4=e_5$, $e_2e_4=e_6$ and 
$e_3e_4 = e_7$, but we are still lacking the relation between
the remaining possible triplets, 
$\{e_1,e_6,e_7\};\{e_2,e_5,e_7\};\{e_3,e_5,e_6\}$
which can be fixed using
\begin{eqnarray}\label{fkf}\begin{array}{c}
e_1 e_6 = e_1 (e_2 e_4) = - (e_1 e_2) e_4 = - e_3 e_4 = - e_7 ,\\
e_2 e_5 = e_2 (e_1 e_4) = - (e_2 e_1) e_4 = + e_3 e_4 = + e_7 ,\\
e_3 e_5 = e_3 (e_1 e_4) = - (e_3 e_1) e_4 = - e_2 e_4 = - e_6 .
\end{array}
\end{eqnarray}
That is all
for octonions.
Going up to $\cal X$, we have the seven octonionic conditions,
and the decomposition
(\ref{ffff}) gives us
$e_1e_8=e_9, e_2e_8=e_{A}, e_3e_8=e_B, e_4e_8=e_C, 
e_5e_8=e_D, e_6e_8= e_E, e_7e_8=e_F$ where
$A=10,B=11,C=12,D=13,E=14$ and $F=15$. The other elements
of the multiplication table may be chosen in analogy with  (\ref{fkf}),
explicitly, the 35 Hexagonionic
triplets N are
\begin{eqnarray}\begin{array}{ccccccc}
(123), &(145), &(246), &(347), &(257), &(176), &(365), \\
(189), &(28A), &(38B), &(48C), &(58D), &(68E), &(78F), \\
(1BA), &(1DC), &(1EF), &(29B), &(2EC), &(2FD), &(3A9), \\
(49D), &(4AE), &(4BF), &(3FC), &(3DE), &(5C9), &(5AF), \\
(5EB), &(6FD), &(6CA), &(6BD), &(79E), &(7DA), &(7CB), \end{array}
\end{eqnarray}
and so on for any generic higher dimensional ``field'' ${\cal F}^n$. 

In general, from some combinatorics, the number of such 
triplets
for  a general ${\cal F}^n$ field is ($n>1$) 
\begin{equation}
N = {\frac{~~~\left( 2^n -1 \right)!~~~}{~\left( 2^n - 3 \right)!~~~ 3!~}} ,
\end{equation}
giving
\begin{eqnarray}
\begin{array}{cccc}
{\cal F}^n&n&~~~~dim~~~~&N\\
{\cal Q}~~~~&2&4&1\\
{\cal O}~~~~&3&8&7\\
{\cal X}~~~~&4&16&35\\
& &and\ so\ on.&\\
\end{array}
\end{eqnarray}
One may notice that for any non-ring division algebra
$\left( {\cal F},\ n>3 \right)$,\  
$N> dim({\cal F}^n)$ except when dim = $\infty$ i.e a functional
Hilbert space with a Cliff(0,$\infty$) structure. Does 
this inequality have  any relation
with the  ring division structure of the
($S^1,S^3,S^7$)  spheres~?! Yes, that is what we are going to show now :
Following, the same translation\footnote{ The translation idea
was given first in \cite{rot2} from $\cal Q$ to ${\cal C}^2$ in the context
of quaternionic quantum mechanics.} idea - projecting our
algebra over $R^{16}$\cite{my1} - any $E_i$ is given by similar
relation as for $\cal Q$ or $\cal O$
\begin{equation}
(E_i)_{\alpha\beta} = \delta_{i\alpha} \delta_{\beta 0} - \delta_{i \beta}
\delta_{\alpha 0} + f_{i\alpha\beta} .
\end{equation}
But contrary to the quaternions and octonions, the Clifford algebra
closes for a subset of these $E_i$'s, namely
\begin{equation}\label{nrd}
\{ E_i , E_j \} = - 2 \delta_{ij}  \quad \mbox{for} \quad i,j,k = 1 \ldots
8.
\end{equation}
Because, we have lost the ring division structure. Also, notice that
(\ref{nrd}) is in agreement with the Clifford algebra
classification\cite{abs}. Following this method,
we can give a simple way to write 
real Clifford algebras over any arbitrary dimensions.

Sometimes, a specific multiplication table may be favoured.
As we are interested in solitons, the existence of a symplectic
structure - related to the bihamiltonian formulation of 
integrable models - should be welcome.
It is known from the Darboux theorem, that locally a symplectic 
structure is given up to a minus sign by 
\begin{equation}
{\cal J}_{dim\times dim} = \left(
\begin{array}{cc}
0& - 1_{\frac{dim}{2}\times \frac{dim}{2}}\\
1_{\frac{dim}{2}\times \frac{dim}{2}}& 0
\end{array} \right) 
, \end{equation}
that fixes the following structure constants
\begin{eqnarray}
& &f_{ \left( {\frac{dim}{2}}    \right) 1 
    \left( {\frac{dim}{2}} +1 \right) } 
= -1, \\
& &f_{ \left( {\frac{dim}{2}}    \right) 2 
    \left( {\frac{dim}{2}} +2 \right) } 
= -1, \\
& & ~~~~~~~\ldots \\
& &f_{ \left( {\frac{dim}{2}}    \right) 
    \left( {\frac{dim}{2}} -1 \right)
    \left(     dim-1          \right) } = -1, 
\end{eqnarray}
which is clearly the decomposition that we have chosen in 
(\ref{a4}) for octonions
\begin{equation}
f_{415} = f_{426} = f_{437} = -1 .
\end{equation}
Generally our symplectic structure
is 
\begin{equation}
\left(1|E_{\left( {\frac{dim}{2}} \right)} 
\right)_{\alpha\beta} = 
\delta_{0\alpha} \delta_{\beta\left( {\frac{dim}{2}} \right)} 
-\delta_{0\beta} \delta_{\alpha\left( {\frac{dim}{2}} \right)} 
- \epsilon_{\alpha\beta\left( {\frac{dim}{2}} \right)}  .
\end{equation} Moreover some other choices may exhibit a relation 
with number theory and Galois fields \cite{dix}. 
It is highly non-trivial how  Clifford algebraic language
can be used to unify many distinct mathematical
notions such as Grassmanian \cite{my2}, complex, quaternionic and symplectic 
structures.

\section{Conclusion}

Octonions have a central in mathematics and they
play a vital role for $D=11$ supergravity compactification.
Understanding their real job in physics is highly
needed, espcially, with the recent string dualities.
In this article, the important message is to use
the associative non Lie algebra
Cliff(0,7) instead of octonions which led us to structure
matrices. Once this is accepted. The road is open.
For example, using this Clifford language, we have
\begin{equation}
E_a E_b = \epsilon_{abcdefg} E_c E_d E_e E_f E_g.
\end{equation}
extending it to $R^8$, we have a natural
8 dimensions Levi-Civita and we may study a self duality
relation for the Reimanian tensor
\begin{equation}
R_{\alpha \beta \gamma \delta } =
\epsilon_{\alpha \beta \gamma \delta \zeta \eta \mu \nu}
R_{\zeta \eta \mu \nu }.
\end{equation}
So, in 2 dimensions, we have a dual $\sigma$ model (constrained
scalar field), in 4 dimensions, a dual YM field (constrained
spin one field) and lastly in 8 dimensions, a dual gravitational
field (constrained spin 2 field) !

Once we write the self duality relation in a certain 
higher dimension, then we can 
recouver the lower cases by trivial dimension reduction.
A very interesting situation happens in odd dimensions,
instead of considering the self duality condition
(\ref{star}), we may generalize the Chern-Simons form
to 7 dimensions as
\begin{equation}\label{ch1}
Ch = \rho_{ijk} (A_i A_j A_k - g A_i \partial_j A_k) ,
\end{equation}
for suitable $g$ to be defined appropriately in correspondence with
$\lambda$.
The first part of (\ref{ch1}) is the natural generalization of the
WZNW term to octonions\footnote{Note that the natural dimension
of an octonionic sigma model is 6 not 2 dimensions as
$\pi_3(S^7) \neq Z$ which may be related to 5-brane.}
whereas the second part
is the octonionic Hopf term. Also,
we may define a vector product and a curl operator over $R^7$ or
any of its subspaces by
\begin{equation}\label{vp}
A_i = \rho_{ijk} A_j A_k
\quad ; \quad B_i = \rho_{ijk} \partial_j A_k .
\end{equation}
The 7 dimensional monopole is just the Clifford self-duality,
making the $A_{1\dots7}$ static
and defining $A_0 = \phi$ our Higgs field. 

We mentioned the  $so(n)$ series embeding but its
extension to any Lie algebra should be clear as
\begin{eqnarray}
su(n) \subset so(2n),\\ 
sp(n) \subset so(4n),
\end{eqnarray}
and having the following interesting topological facts (taking into
account Bott periodicity) 
\begin{eqnarray}
\pi_{2^n-1}(so(2^n-1)) &=& Z, \\
\pi_{2^n-1}(so(2^n)) &=& Z \oplus Z, \\
\pi_{2^n-1}(su(2^n)) &=& Z , \\
\pi_{2^n-1}(sp(2^{n-1})) &=& Z ,
\end{eqnarray}
also, noticing that
\begin{equation}
\pi_{15}(H) = Z \quad for \quad H=F_4,E_6,E_7,E_8 , 
\end{equation}
which hopefully may be related to the $E_8\times E_8$
string solitons \cite{pp3,gun1,Hst,duf1,duf2}.

As physics speaks mathematics, we tried our best to adopt
Dirac's point of view: ``one can
generalize his physics by generalizing his mathematics'', a line
of attack that always proved to be useful.

\vspace{3cm}
Just before the submission of this work, we received a preprint\cite{last}
where a Cliff(0,7) is also used instead of octonions. But it is
clear that we addressed different questions related to the 8 dimensional
instanton.

\vspace{3.5 cm}

\section*{Acknowledgements}

I would like to thank many people who gave me from their
knowledge.
Especially, 
P.~Rotelli and
G.~Thompson. Also, I am grateful to M.~Boiti, O.~Pashaev, F.~Pempinelli
and G.~Soliani for many fruitful discussions about higher
dimensions solitons and to G.~De~Cecco for some topological useful
informations.
Last, It is a pleasure to acknowledge  Prof. A.~Zichichi 
and the ICSC--World Laboratory for financial
support.

\vspace{3cm}

\vspace{3cm}
\begin{flushright}
{\bf
~~~~~Khaled Abdel-Khalek~~~~~~~~~~~~~ \\
~~~~~Dipartimento di Fisica \&~~~~~~~~~~ \\
Istituto Nazionale di Fisica Nucleare\\
~~~~~- Universit\`a di Lecce -~~~~~~~~~~~\\
~~~~~- Lecce, 73100, Italy -~~~~~~~~~~~}
\end{flushright}
\end{document}